\documentclass{book}
\usepackage{piers}
\begin{document}

\title{Chaotic Communication with Robust Hyperbolic\\ Transmitter and Receiver}
\maketitle

\author      {O. B. Isaeva}
\affiliation {Department of Nonlinear Processes, Chernyshevsky Saratov State University}
\address     {}
\city        {Saratov}
\postalcode  {}
\country     {Russia}
\phone       {+7(8452)278685}    
\fax         {233445}    
\email       {email@email.com}  
\misc        { }  
\nomakeauthor

\author      {A. Yu. Jalnine}
\affiliation {Saratov Branch of Kotelnikov's Institute of Radio-Engineering and Electronics of RAS}
\address     {}
\city        {Saratov}
\postalcode  {}
\country     {Russia}
\phone       {345566}    
\fax         {233445}    
\email       {jalnine@rambler.ru}  
\misc        { }  
\nomakeauthor

\begin{authors}
{\bf O. B. Isaeva}$^{1,2}$, {\bf A. Yu. Jalnine}$^{2}$, {\bf and S. P. Kuznetsov}$^{2,3}$\\
\medskip
$^{1}$Department of Nonlinear Processes, Chernyshevsky Saratov State University, Russia\\
$^{2}$Saratov Branch of Kotelnikov's Institute of Radio-Engineering and Electronics of RAS, Russia\\
$^{3}$Institute of Mathematics, Information Technologies and Physics, Udmurt State University, Russia
\end{authors}

\begin{paper}

\begin{piersabstract}
New chaos-based communication schemes for transmission of analog and digital information are suggested. The carrier signal is produced by chaotic generator having well-defined oscillation phases at least on short time intervals. For data extraction, a special procedure of phase detection for chaotic signals is developed. An efficiency of the scheme is demonstrated by examples of coupled generators of structurally stable (robust) chaos with hyperbolic attractors of Smale-Williams and Arnold's ``cat map'' types. Robustness of the transmitter and the receiver makes it possible to transmit information in presence of parameter mismatch or additive noise in the communication channel.
\end{piersabstract}

\psection{Introduction}
In recent years much attention of researchers has been paid to the problem of using chaotic dynamical systems in different communication schemes [1-3]. Hypothetically, a chaotic signal as an information carrier has some advantages compared with regular ones. First, it has larger information capacity and better noise immunity at moderate signal intensities due to the broadbandness. Second, a fundamental phenomenon of chaotic synchronization [4] can be used effectively: when the transmitter and receiver are coupled in appropriate way, completely synchronous although chaotic behavior arises without any additional controlling scheme. Third, such communication schemes are ``secure'' to some extent, since it is necessary to restore the dynamics of the chaotic transmitter for decoding of the information.

As known, the mostly perspective method for transmission of analog information consists in ``nonlinear mixing'' of information signal into the chaotic dynamics of the transmitting generator [5,6]. In the communication systems embodying this principle the information signal affects upon the transmitting generator and controls its chaotic dynamics via a feedback loop. The chaotic signal generated in this way then combines with the information signal transmitted in the communication channel. Affecting upon the receiver, the composite signal provides synchronous chaotic response, which can be extracted through the open feedback loop of the receiver and then used for reproducing information from the transmitted composite signal.

However, one should note two problems, which obstruct direct realization of this scheme in practice. The first problem is that in majority of such schemes the composite signal being transmitted in the communication channel appears to be a superposition of a weak informational and a strong chaotic component, which are subtracted during the detection process; it results in an ineffective use of energy and in loss of noise immunity. The second problem consists in a high sensitivity of chaotic synchronization to parameter mismatch and to internal noises of the subsystems; that hinders appearance of the completely identical chaotic response in the receiver necessary for the information detection. The purpose of the present work is to indicate some ways for overcoming these two problems.

We suggest modification of the ``nonlinear mixing'' communication scheme. The novelty of the modification consists in (i) usage of {\it phase modulation} of the pseudo-harmonic carrier signal with a chaotically floating phase for the information transmission and (ii) exploiting generators of {\it robust chaos} [7] as the transmitter and the receiver. As in the traditional case of a simple harmonic carrier, usage of the phase modulation principle must augment stability of the scheme with respect to additive noises, which distort the amplitude profile of the signal in the channel. On the other hand, using robust generators of hyperbolic chaos [7-10] as transmitter and receiver resolves the problem of sensitivity of chaos synchronization regimes to parameter mismatches and internal noises, since the hyperbolic strange attractor is weakly sensitive to perturbations of equations of the system [11].

\psection{Chaotic phase modulation and detection procedure}
The suggested procedure of phase modulation of chaotic carrier signals is rather similar to that used for a simple harmonic carrier. Let us have a chaotic signal of pseudo-harmonic type: $y(t)=A\cos(\omega t+\varphi)$ , where $\omega(t)$ and $A(t)$ are chaotic functions of time, which describe the slowly changing amplitude and phase, respectively ($\dot{\varphi}\ll\omega$,$\dot{A}/A\ll\omega$). We need to add information $\xi(t)$ (where $\xi\ll 1$ and $\dot{\xi}\ll \omega$) to this signal via the modulation of the chaotic phase: $\varphi\rightarrow \varphi+\xi$ to obtain new signal $s(t)=A\cos(\omega t+\varphi+\xi)$. In order to do this, we set
\begin{displaymath}
\begin{array}{lll}
s(t)=A\cos(\omega t+\varphi(t)+\xi(t))\approx A\cos(\omega t+\varphi(t))-A\xi(t)\sin(\omega t+\varphi(t))\approx \\
\approx A\cos(\omega t+\varphi(t))-A\xi(t)\cos(\omega(t-\pi/2\omega)+\varphi(t-\pi/2\omega)) = y(t)-y(t-\pi/2\omega)\xi(t).
\end{array}
\end{displaymath}
Thus, the phase-modulated signal in the communication channel will be a superposition of the original chaotic signal with itself, but with a quarter-period delay and multiplied by the information signal
\begin{equation}
s(t)\approx y(t)-y(t-\pi/2\omega)\xi(t).
\label{eq1}
\end{equation}
For extraction of information, let us use the modified procedure of the phase detection. It consists in composition of the transmitted phase-modulated signal with a reference signal to provide its transformation to an amplitude-modulated signal, with further amplitude detection. In our case, the reference signal is of the form:
\begin{equation}
s_{ref}(t)= -y(t)-y(t-\pi/2\omega)\xi(t).
\label{eq2}
\end{equation}
Summarizing, we obtain
\begin{displaymath}
\tilde{s}(t)=s+s_{ref}=-(1+\xi(t))y(t-\pi/2\omega).
\end{displaymath}
From the last expression one can see that $\tilde{s}(t)$ is a chaotic signal modulated by the informational signal $\xi(t)$. Since $|\dot{\xi}|\ll\omega$, a variation of $\xi$ during one quasi-period $T=2\pi/\omega$ is minor. Hence, one can average the last expression over a single interval $T$ to obtain:
\begin{equation}
\xi'(t)\approx\frac{<|\tilde{s}(t)|>}{<|y(t-\pi/2\omega)|>}-1,
\label{eq3}
\end{equation}
where $<f(t)>=\int^{t+\pi/\omega}_{t-\pi/\omega}f(\tau)d\tau$. Thus, the procedures of phase modulation and detection for the chaotic signals of quasi-harmonic type are quite analogous to those used traditionally for simple harmonic signals.

\psection{Communication with coupled R\"{o}ssler oscillators}
As illustration of the procedure we suggest to consider a communication scheme based on coupled modified R\"{o}ssler oscillators:
\begin{equation}
\dot{x}=-\omega y-z, \quad \dot{y}=\omega x+ay-z, \quad \dot{z}=r+z(x+y-c).
\label{eq4}
\end{equation}
The sense of the modification consists in increase of the phase dispersion compared with the classical equations. This system is not regarded as a ``robust'' one. However, for parameter values $\omega=1.0$, $a=0.2$, $r=0.65$, $c=11$ it manifests a phase-coherent chaotic self-oscillatory regime, which corresponds to a spiral attractor in the phase space [12], so that it is convenient to use it for illustration. The attractor is characterized by a set of Lyapunov exponents $\lambda_1\approx0.071$, $\lambda_2\approx0.0$, $\lambda_3\approx-10.7$, by an average angle frequency $\omega_0=2\pi/T\approx 1.019$ and by the effective phase dispersion coefficient $D\approx7.8\times 10^{-4}$. The projection of the phase portrait of the attractor in coordinates $(x,y)$ is shown in Fig.1(a). The fragments of time dependencies for $x(t)$ and $y(t)$ are presented in Fig.1(b). From these figures one can see that the time dynamics of $x(t)$ coincides with the quarter-period shifted $y(t)$ well, as it should occur for a case of quasi-harmonic oscillations. Hence, in the phase modulation procedure~(\ref{eq1}) we can use a substitution ``$-y(t-\pi/2\omega_0\rightarrow x(t))$''. Let us consider a pair of unidirectionally coupled oscillators~(\ref{eq4}), where the driving (D) subsystem operates as a transmitter, while the response (R) subsystem plays a role of the receiver:
\begin{equation}
D:\left\{ \begin{array}{ll}
\dot{x}_d=-\omega y_d-z_d, \\ \dot{y}_d=\omega x_d+a s(t)-z_d, \\ \dot{z}_d=r+z_d(x_d+y_d-c),
\end{array} \right.
\quad
R:\left\{ \begin{array}{ll}
\dot{x}_r=-\omega y_r-z_r, \\ \dot{y}_r=\omega x_r+a s(t)-z_r, \\ \dot{z}_r=r+z_r(x_r+y_r-c),
\end{array} \right.
\label{eq5}
\end{equation}
where $s(t)=y_d+x_d\xi(t)$.

\begin{figure}
\begin{center}
\resizebox{0.75\columnwidth}{!}{\includegraphics{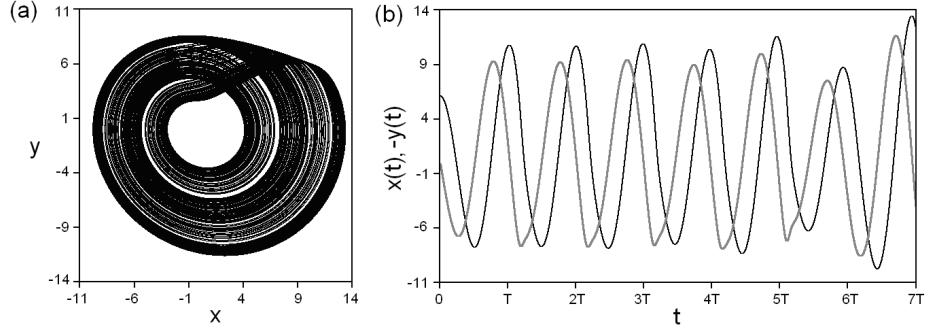} }
\end{center}
\caption{(a) Projection of the phase portrait and (b) time dependencies $x(t)$ (black) and $y(t)$ (grey) of the system~(\ref{eq4}).}
\label{fig:f1}
\end{figure}

When the informational signal is absent ($\xi(t)=0$), the coupling of the systems via the variable $y$ in the second equation gives rise to synchronous chaotic response [5], so that the dynamics of the corresponding variables become completely identical ($x_d=x_r$, $y_d=y_r$, $z_d=z_r$). Such synchronization is characterized by negative conditional Lyapunov exponents associated with the response system: $\lambda_1^r\approx-0.025$, $\lambda_2^r\approx-0.109$, $\lambda_3^r\approx-10.7$. In presence of a small information signal, which does not destroy the dynamics of the system, the synchronous regime also retains. Since $x_d=x_r$ and $y_d=y_r$, the expression for the reference signal can be written as: $s_{ref}(t)=y_r+x_r$. Then, the amplitude-modulated signal is:
\begin{equation}
\tilde{s}(t)=s(t)+s_{ref}(t)=-(1+\xi(t))x_r(t),
\label{eq6}
\end{equation}
and the detected informational signal $\xi'(t)$ can be recovered as:
\begin{equation}
\xi'(t)=\frac{<|\tilde{s}(t)|>}{<|x_r(t)|>}-1,
\label{eq7}
\end{equation}

For the computer simulation of the information transmission and detection, let us chose $\xi(t)$ in the form of the complex broadband quasiperiodic signal:
\begin{equation}
\xi(t)=A_1\cos(\Omega_1 t+A_2\cos\Omega_2 t),
\label{eq8}
\end{equation}
where the ratio $\Omega_1/\Omega_2$ is irrational. Consider phase dynamics of the variable $s(t)$ in the communication channel for the cases of absence and presence of the information transmission. At each moment the phase can be defined via differentiation of the transmitted signal as $\varphi_s=\arg(s-i\dot{s}/\omega_0)$. When the information signal is absent, this phase coincides with the phase of the variable $y$: $\varphi_y=\arg(y-i\dot{y}/\omega_0)$. In Fig.2(a) the time dependence of the phase $\varphi_y(t)$ is shown for $\xi(t)=0$. The next plot (Fig.2(b)) shows the plot for $\varphi_s(t)$ in presence of the informational signal~(\ref{eq8}) with $\Omega_1=0.1\omega$, $\Omega_2=\Omega_1(\sqrt{5}-1)/2$, $A_1=0.1$, $A_2=0.5$. Comparing these plots, one can see that presence of a small informational signal does not produce a notable effect on dynamics of the chaotic phase; it allows us to say about masking of the transmitted information.
\begin{figure}
\begin{center}
\resizebox{0.75\columnwidth}{!}{\includegraphics{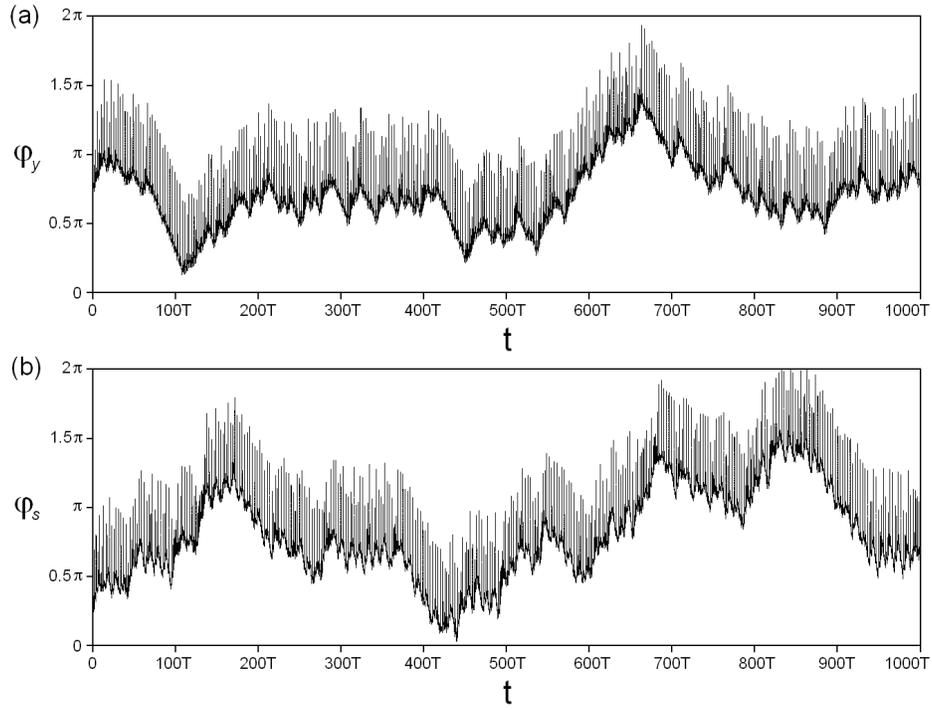} }
\end{center}
\caption{The dynamics of the chaotic phase of the signal in communication channel (a) in absence and (b) in presence of the information transmission.}
\label{fig:f2}
\end{figure}
In order to obtain approximate ``transcription'' of the informational signal, consider the phase difference: $\Delta\varphi=\varphi_s-\varphi_y$. The respective plot is shown in the Fig.3(a) (parameters of the information signal are the same as in Fig.2(b)). Here the time dependence $\Delta\varphi(t)$ is shown in grey, and the original information signal $\xi(t)$ is shown in black. From this comparison one can easily see that the ``differential'' approximation contains high-frequency oscillations. Hence, for exact detection, two terms must be fulfilled: (i) the condition $\dot{xi}\ll \omega_0$ must hold, and (ii) the detection procedure via the formulas~(\ref{eq6}) and~(\ref{eq7}) must be implemented.

Results of computer simulation of the transmission and detection of information signal~(\ref{eq8}) with $\Omega_1=10^{-3}\omega$, $\Omega_2=\Omega_1(\sqrt{5}-1)/2$, $A_1=0.1$, $A_2=0.5$ are illustrated by Fig.3(b). A fragment of the amplitude-modulated signal $\tilde{s}(t)$ obtained in accordance with~(\ref{eq6}) is shown in grey, while the informational signal $\xi'(t)$ reconstructed as the amplitude envelope is shown in black. The detected signal entirely coincides with the original one $\xi(t)$, just as expected.
\begin{figure}
\begin{center}
\resizebox{0.75\columnwidth}{!}{\includegraphics{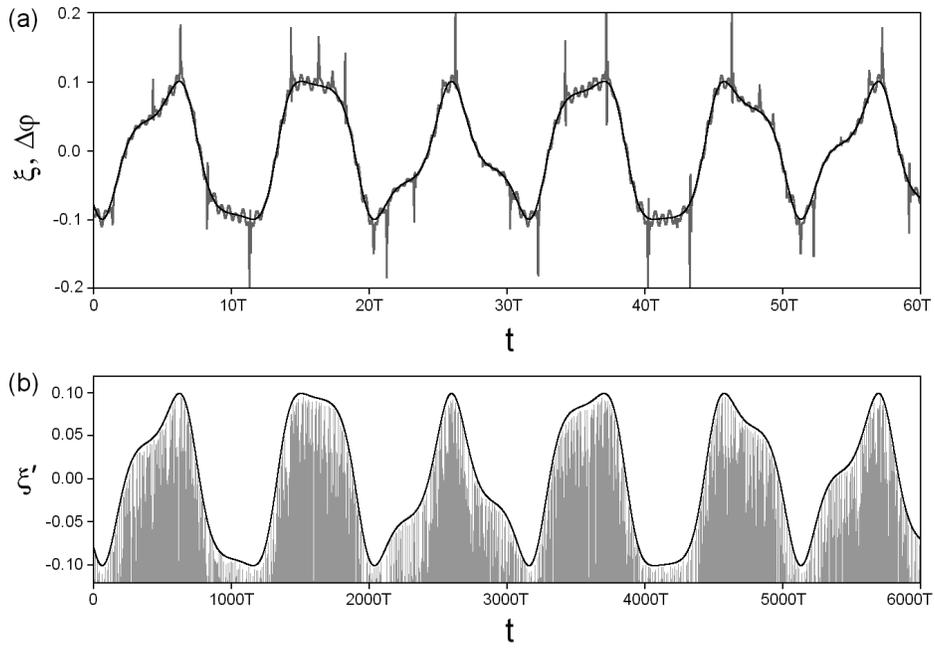} }
\end{center}
\caption{Extraction of information from (a) the phase of the transmitted signal via estimation of the phase difference $\Delta\varphi=\varphi_s-\varphi_y$, and (b) via the procedure of chaotic phase detection. }
\label{fig:f3}
\end{figure}

\psection{Communication with coupled generators of hyperbolic chaos}
For next illustration of operability of our communication scheme, let us apply a generator of robust hyperbolic chaos suggested first in the work [8] and realized later as an electronic device [9]. The dynamical equations are
\begin{equation}
\begin{array}{ll}
\ddot{x}-(A\cos{\omega_0 t/N}-x^2)\dot{x}+\omega_0^2x=\varepsilon y\cos{\omega_0 t}, \\
\ddot{y}-(-A\cos{\omega_0 t/N}-y^2)\dot{y}+(2\omega_0)^2 y=\varepsilon x^2.
\end{array}
\label{eq9}
\end{equation}
For the parameter values $A=3.0$, $\varepsilon=0.5$, $\omega_0=2\pi$, $N=10$ the system~(\ref{eq9}) demonstrates hyperbolic strange attractor of Smale-Williams type, which is characterized by a set of Lyapunov exponents $\lambda_1\approx 0.069$, $\lambda_2\approx -0.35$, $\lambda_3\approx -0.59$, $\lambda_4\approx-0.81$ and by dynamics of the phase $\varphi=\arg(x-i\dot{x}/\omega_0)$ approximately given by the Bernoulli map: $\varphi'\approx 2\varphi+Const$.

A pair of unidirectionally coupled systems~(\ref{eq9}) can be used as transmitter and receiver in the communication scheme. Namely, we set:
\begin{equation}
\left\{
\begin{array}{ll}
\ddot{x}_d-(A\cos{\omega_0 t/N}-x_d^2)\dot{x}_d+\omega_0^2 x_d=\varepsilon s(t), \\
\ddot{y}_d+(A\cos{\omega_0 t/N}+y_d^2)\dot{y}_d+4\omega_0^2 y_d=\varepsilon x_d^2,
\end{array} \right.
\left\{
\begin{array}{ll}
\ddot{x}_r-(A\cos{\omega_0 t/N}-x_r^2)\dot{x}_r+\omega_0^2 x_r=\varepsilon s(t), \\
\ddot{y}_r+(A\cos{\omega_0 t/N}+y_r^2)\dot{y}_r+4\omega_0^2 y_r=\varepsilon x_r^2,
\end{array} \right.
\label{eq10}
\end{equation}
The coupling between the transmitter and receiver is provided by a common term in the first equation of each subsystem:
\begin{equation}
s(t)=y_d\cos(\omega_0 t+\xi(t)),
\label{eq11}
\end{equation}
where $\xi(t)$ is the information signal to be transmitted. In the absence of information mixing ($\xi(t)=0$), the appearing stable regime of complete synchronization is characterized by a set of $4$ conditional Lyapunov exponents associated with the response system: $\lambda_1^r\approx -0.275$, $\lambda_2^r\approx -0.278$, $\lambda_3^r\approx -0.557$, $\lambda_4^r\approx-0.564$. Since the chaotic regime is structurally stable, it retains after introducing a (relatively small) information effect upon the transmitting system; the complete synchronization of subsystems also holds ($x_d=x_r$, $y_d=y_r$). Therefore, detecting of the information signal can be produced via the scheme described above. First, we add the reference signal $s_{ref}(t)=-y_r\cos{\omega_0 t}-y_r\sin{\omega_0 t}$ to the transmitted signal $s(t)$; after that, the obtained amplitude-modulated signal $\tilde{s}(t)=-y_r(1+\xi(t))\sin{\omega_0 t}$ is detected via extraction of the amplitude envelope (averaging):
\begin{equation}
\xi'(t)=\frac{<|\tilde{s}(t)|>}{<|y_r(t)\sin{\omega_0 t}|>}-1.
\label{eq12}
\end{equation}

Results of the computer simulation of the transmission and detection of the complex broadband analog signal are illustrated in Fig.4. The original information signal of the form~(\ref{eq8}) with  $\Omega_1=10^{-3}\omega$, $\Omega_2=\Omega_1(\sqrt{5}-1)/2$, $A_1=0.1$, $A_2=0.5$ (Fig.4(a)) affects upon the transmitter; the phase-modulated signal of the form~(\ref{eq11}) is forwarded to the communication channel, where it has an appearance of the pulses with chaotic phase dynamics (Fig.4(b)). After the channel passage the signal undergoes conversion to the amplitude-modulated signal with further detection of the amplitude envelope. Finally, the obtained information signal is observed coinciding with the original one (Fig.4(c)).
\begin{figure}
\begin{center}
\resizebox{0.75\columnwidth}{!}{\includegraphics{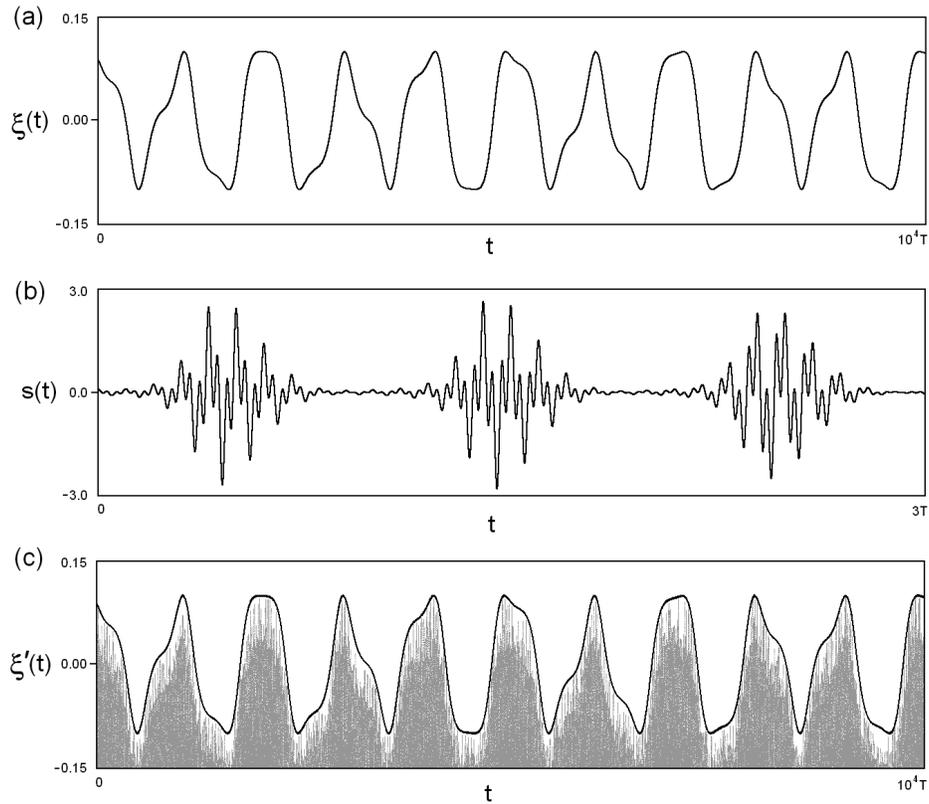} }
\end{center}
\caption{Illustration of the procedure of information transmission on the basis of the system~(\ref{eq10}): (a) the information signal to be transmitted, (b) the sequence of phase-modulated chaotic pulses in the communication channel, (c) the amplitude envelop extraction for detecting of information.}
\label{fig:f4}
\end{figure}

\psection{Transmission of digital information via robust chaotic communication}
Previously, we discussed transmission of analog information via chaotic phase modulation. In this section we consider digital communication on basis of hyperbolic chaos generators. The first scheme is associated with equations~(\ref{eq10}); it is represented as a flowchart in Fig.5. A sample of information to be transmitted is a gray-scale image with 256 gradations. Thus, we need an 8 bit alphabet for transmission of a single pixel. The information function $\xi=I(t)$ will be defined as a step function $0<I(t)<1$, with the step length $\tau=10\pi N/\omega_0$ (i.e., 5 periods of parameter modulation in~(\ref{eq9}) or~(\ref{eq10}), and with the step height corresponding to the intensity of the pixel color.
\begin{figure}
\begin{center}
\resizebox{1.0\columnwidth}{!}{\includegraphics{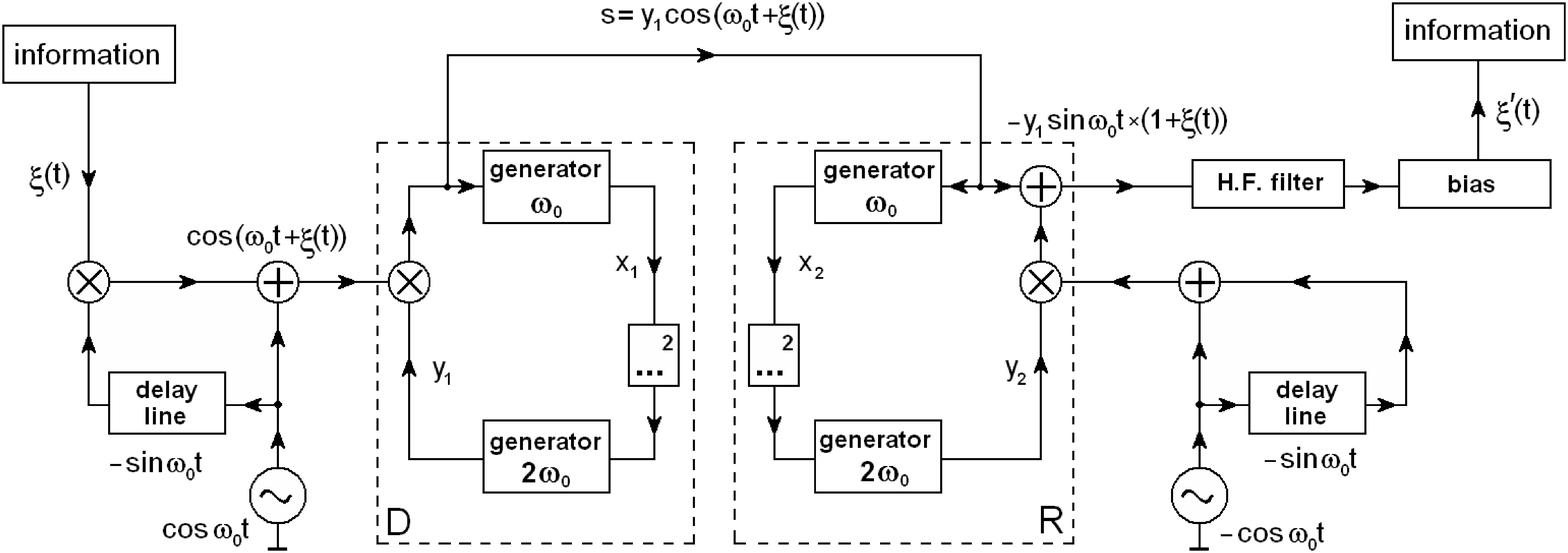} }
\end{center}
\caption{Flowchart of the system~(\ref{eq10}).}
\label{fig:f5}
\end{figure}
The operability of the scheme was checked for cases of strongly identical and slightly non-identical transmitter and receiver. Figure 6 shows the original image to be transmitted (a), the ``image'' obtained by straightforward processing the signal in the communication channel after chaotic phase modulation (b), and finally, the restored image after the chaotic phase detection procedure applied (c). The images in the Figs.6(a) and (c) are completely identical, just as expected.

The next two panels of Fig.6 show the decoded images in presence of parameter mismatch between the transmitter and the receiver: $\Delta A=0.1$  (d) and $\Delta\omega_0=10^{-6}$ (e). Note, that sufficiently large mismatches of the internal parameter A of the hyperbolic chaos generator almost do not distort the transmitted image, while a very small lack of frequency coincidence results in the full loss of the image quality. However, such effect can be avoided by adjustment of the external driving parameters of the transmitter.
\begin{figure}
\begin{center}
\resizebox{1.0\columnwidth}{!}{\includegraphics{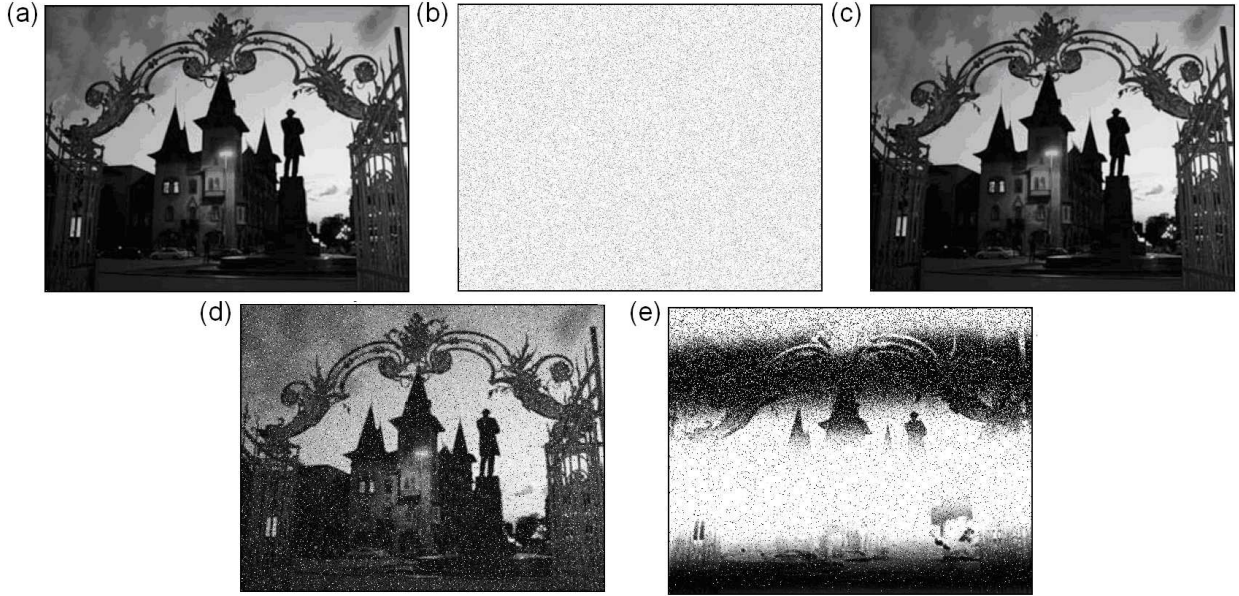} }
\end{center}
\caption{The illustration of digital image transmission via the system~(\ref{eq10}). See explanations in the text. }
\label{fig:f6}
\end{figure}
Another example of the digital communication scheme based on robust chaos generators engages the oscillatory system which reproduces the dynamics of the conservative Arnold's ``cat map'' in the stroboscopic section of the flow [10]. A single underlying system consists of four van der Pol oscillators with modulated parameters of excitation. A pair of such systems with unidirectional coupling can operate as the transmitter/receiver of the communication scheme in accordance with the equations:
\begin{equation}
\left\{
\begin{array}{ll}
\ddot{x}_d-(A\cos{\omega_0 t/N}-x_d^2)\dot{x}_d+\omega_0^2 x_d=\varepsilon s(t), \\
\ddot{y}_d-(A\cos{\omega_0 t/N}-y_d^2)\dot{y}_d+\omega_0^2 y_d=\varepsilon w_d, \\
\ddot{z}_d+(A\cos{\omega_0 t/N}+z_d^2)\dot{z}_d+4\omega_0^2 z_d=\varepsilon x_d y_d, \\
\ddot{w}_d+(A\cos{\omega_0 t/N}+w_d^2)\dot{w}_d+\omega_0^2 z_d=\varepsilon x_d,
\end{array} \right.
\left\{
\begin{array}{ll}
\ddot{x}_r-(A\cos{\omega_0 t/N}-x_r^2)\dot{x}_r+\omega_0^2 x_r=\varepsilon s(t), \\
\ddot{y}_r-(A\cos{\omega_0 t/N}-y_r^2)\dot{y}_r+\omega_0^2 y_r=\varepsilon w_r, \\
\ddot{z}_r+(A\cos{\omega_0 t/N}+z_r^2)\dot{z}_r+4\omega_0^2 z_r=\varepsilon x_r y_r, \\
\ddot{w}_r+(A\cos{\omega_0 t/N}+w_r^2)\dot{w}_r+\omega_0^2 z_r=\varepsilon x_r,
\end{array} \right.
\label{eq13}
\end{equation}
where $s(t)=z_d\cos(\omega_0 t+\xi(t))$. The parameter values are chosen as in [10]: $A=2.0$, $\varepsilon=0.4$, $\varepsilon=2\pi$, $N=20$. From the viewpoint of confidentiality, such scheme has an advantage due to higher dimension of the phase space; that makes difficulties for the decoding of the potentially intercepted signal. Figure 7 illustrates the transmission of a grey-scaled image (panel (a)) encoded with a step function (as previously, the function has 256 discrete values, the time length of a single step is $5T$, where $T=2\pi N/\omega_0$) in the presence of the parameter mismatch $\Delta A=0.1$ (b) and frequency mismatch $\Delta \omega_0=10^{-6}$ (c). The scheme demonstrates rather steady transmission quality in respect to perturbation of the internal parameter $A$ and rather high sensitivity to the mismatch of the frequency of external modulation $\omega_0$.
\begin{figure}
\begin{center}
\resizebox{1.0\columnwidth}{!}{\includegraphics{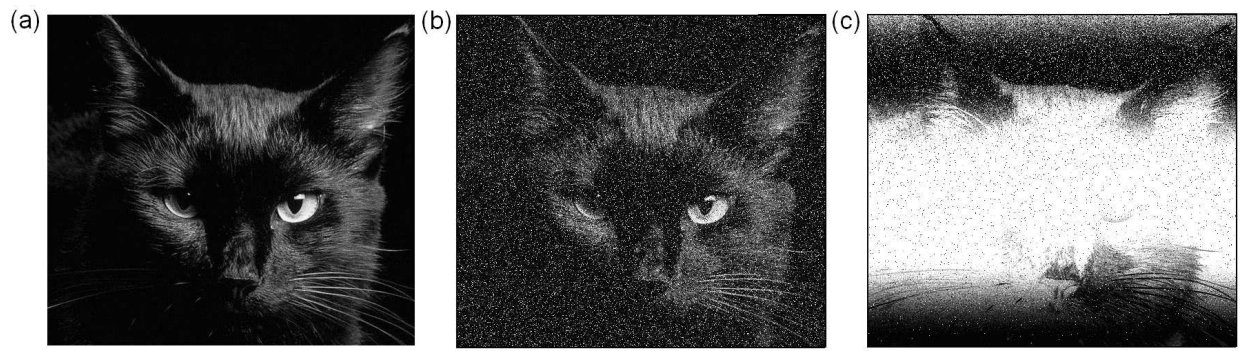} }
\end{center}
\caption{The illustration of digital image transmission via the system~(\ref{eq13}). See explanations in the text.}
\label{fig:f7}
\end{figure}

\psection{Conclusion}
The suggested schemes implement simultaneously several relevant concepts in the chaotic communication: (i) retaining principles of broadbandness of the chaotic carrier signal, self-synchronization of the transmitter and receiver, conditional ``confidentiality'' of information transmission; (ii) exploiting the phase modulation principle, the design of chaotic phase modulation and detection, which uses the complete power of the chaotic carrier (energy efficiency, noise immunity); (iii) usage of robust (hyperbolic) chaos generators as the transmitter and receiver. Application of these schemes may include development of devices of broadband wireless analog and digital communication, which operate at shot distances and do not require allocation of special frequency band because of the intrinsic low spectral power density.

\ack
The work is performed under partial support of RFBR grant No. 16-02-00135.

\end{paper}

\end{document}